\documentclass[a4paper]{article}  %%% Necessary
\usepackage{indentfirst}  %% Text feature after captions
\usepackage{bm}    %%%%%  for mathematical symbols
\usepackage{graphicx}  %%%% Insert eps file using the 1st method, cite this package
\usepackage{url}
\usepackage{fancyhdr}
\usepackage{mathtools}
\usepackage{enumitem}
\usepackage{subfigure}
\usepackage{tabularx}
\usepackage{multirow}
\usepackage[table,xcdraw]{xcolor}
\usepackage{url}
\usepackage{fancyhdr}
\usepackage{geometry}
\begin{document}    %% Paper start here
%% replace the XXX with your paper number
%-------------------  First Head  -----------------------------------------
\thispagestyle{empty} \vspace*{0.8cm}\hbox
to\textwidth{\vbox{\hfill\noindent \\ \textit{Proceedings of the 8th International Conference on Pedestrian and Evacuation Dynamics (PED2016)\\
Hefei, China - Oct 17 -- 21, 2016\\
Paper No. 42}
\hfill}}
\par\noindent\rule[3mm]{\textwidth}{0.2pt}\hspace*{-\textwidth}\noindent
\rule[2.5mm]{\textwidth}{0.2pt}

%=================== Text begin here =============================================
%% Paper title
\begin{center}
\LARGE\bf Towards Modelling Pedestrian-Vehicle Interactions: Empirical Study on Urban Unsignalized Intersection
\end{center}

%%Authors
\begin{center}
\rm Andrea Gorrini$^{1}$\footnote{Corresponding author.} \ Giuseppe Vizzari${^1}$ and Stefania Bandini$^{1,2}$
\end{center}

\begin{center}
\begin{small} \sl
${}^{\rm 1}$ Complex Systems \& Artificial Intelligence research center, University of Milano-Bicocca \\Viale Sarca 336 - Building U14, 20126 Milano (ITALY) \\   %%%% Address 1 and 2)
andrea.gorrini@unimib.it; giuseppe.vizzari@disco.unimib.it; stefania.bandini@disco.unimib.it \\
${}^{\rm 2}$ Research Center for Advanced Science and Technology, The University of Tokyo \\ 4-6-1 Komaba, Meguro-ku, Tokyo 153-8904 (JAPAN) \\   %%
\end{small}
\end{center}
\vspace*{2mm}

\begin{center}
\begin{minipage}{15.5cm}
\parindent 20pt\small
%%%% Abstract starts here
\noindent\textbf{Abstract -} The modelling and simulation of the interaction among vehicles and pedestrians during cross-walking is an open challenge for both research and practical computational solutions supporting urban/traffic decision makers and managers. The social cost of pedestrians\rq\ risky behaviour pushes the development of a new generation of computational models integrating analytical knowledge, data and experience about the complex dynamics occurring in pedestrian/vehicle interactions, which are not completely understood despite recent efforts. This paper presents the results of a significant data gathering campaign realised at an unsignalized zebra crossing. The selected area of the city of Milan (Italy) is characterised by a significant presence of elderly inhabitants and pedestrian-vehicle risky interactions, testified by a high number of accidents involving pedestrians in the
past years. The results concern the analysis of: (\emph{i}) vehicular and pedestrian traffic volumes; (\emph{ii}) level of service; (\emph{iii}) pedestrian-vehicle interactions, considering the impact of ageing on crossing behaviour. Results showed that the phenomenon is characterised by three main phases: approaching, appraising (evaluation of the distance and speed of oncoming vehicles) and crossing. The final objective of the research is to support the development of a microscopic agent-based tool for simulating pedestrian behaviour at unsignalized crosswalks, focusing on the specific needs of the elderly pedestrians. 
\end{minipage}
\end{center}

\begin{center}
\begin{minipage}{15.5cm}
\begin{minipage}[t]{2.3cm}{\bf Keywords:}\end{minipage}
\begin{minipage}[t]{13.1cm}
%%%%% Keywords start here
Pedestrian, Crossing behaviour, Ageing, Observation
\end{minipage}\par\vglue8pt
\end{minipage}
\end{center}

\section{Introduction} 

The role of advanced computer-based systems for the simulation of vehicular traffic is a consolidated field of research and application that produced results whose level of maturity led to a significant impact on the activity of traffic engineers and planners in the design of efficient transportation networks. Several successful models for the simulation of different aspects of vehicular traffic have been developed and applied: see \cite{nagel2003still} for a review of different approaches, which include both cellular automata discrete models \cite{wang2006modelling} and continuous ones like car-following models \cite{kesting2007general}. 

In parallel, the micro-simulation of pedestrian circulation dynamics have also emerged and affirmed as supports to the assessment of the comfort and safety of urban crowded facilities in case of both ordinary and emergency operations. In analogy with vehicular traffic, the simulation of pedestrian dynamics has been tackled by different approaches, considering both continuous models such as the social force \cite{helbing1995social} and discrete ones based on the floor field method, such as cellular automata \cite{burstedde2001simulation}.

Whereas separately the above mentioned simulation approaches have produced a significant impact, efforts characterised by an integrated investigation on the simultaneous presence of vehicular traffic and pedestrian dynamics are not as frequent or advanced as isolated vehicular traffic and pedestrian models. With the exception of \cite{helbing2005analytical,zeng2014application}, most efforts in this direction are relatively recent or they just analyse simple scenarios not even validated against real data. 

Although pedestrian/vehicle interactions have been empirically studied by different disciplines and methodologies (e.g., transportation engineering, traffic psychology, safety and security science), few attempts towards the validation of models against empirical evidences have been performed. In particular, the design and execution of controlled experiments about the phenomena are limited by several practical and ethical issues about the safety of participants involved in the study. For this reason, the most consolidated methodology in this field is represented by the execution of unobtrusive observations on urban intersections, focused on both aggregated results of traffic volumes and microscopic behavioural indicators.

From a general point of view, observations \cite{gifford} allow to unobtrusively collect empirical data about human behaviour (subjects are not aware to be observed), considering both the physical and social features of the environment in which the subjects are situated. Compared to experiments, this method is characterised by the possibility to exert a limited amount of control over the variables and the environment in which the study takes place, but also by a higher possibility to generalise results (ecological validity).

In this context, the technical report presented in the Highway Capacity Manual \cite{milazzo1999quality} is the most extended study about pedestrian/vehicle interactions, aiming at describing general design standards for enhancing the comfort and safety of both drivers and pedestrians as they travel and walk through urban  intersections or roadway segments (e.g., efficient vehicular transport network, pedestrian exposure to risky crossing). Other relevant observational studies present in the literature more specifically investigated the relevant elements which characterise driving and crossing behaviours. In particular, the results presented in \cite{varhelyi1998drivers} highlighted the impact of traffic volumes and quality of infrastructures on drivers\rq\ compliance with crossing pedestrians, and the effect of perceptive/attentional skills on the adaptation of speed to avoid collisions in time. In analogy with vehicular dynamics, traffic volumes and cross-walk surface conditions were found to significantly influence pedestrians\rq\ crossing decisions \cite{perumal2014study}. More in details, crossing behaviour has been empirically defined as determined by the efficient interaction among locomotion capability \cite{sisiopiku2003pedestrian}, perceptive and attentional skills in evaluating the distance away and the speed of approaching vehicles \cite{hamed2001analysis,sun2015estimation} and individual attitude and motivation towards hazardous situation \cite{evans1998understanding}. 

An innovative and promising methodology in this field of study is represented by the recent possibility to test and measure drivers and pedestrian behaviours by using immersive virtual reality technologies (e.g., image-generation and projection system, virtual reality head-mounted displays). Unlike observational approaches, this method gives the possibility to systematically measure the occurrence and persistence of certain behavioural indicators in a controlled and safe environment, focusing on either driving \cite{roenker2003speed} or crossing behaviours \cite{lobjois2007age}.

All the research efforts in this field are driven by the necessity to develop advanced and sustainable transportation strategies to contrast the social costs of pedestrians\rq\ injury and death due to car accidents \cite{world2013global}. Although the design of efficient, accessible and safe infrastructures is the most important requirement to guarantee the security of the citizens in urban intersections, human factors play a determinant role in the complex interaction among vehicles and pedestrians, also considering the specific needs of vulnerable pedestrians with limited mobility, such as the elderlies. In particular, the examination of the variables which determinate pedestrians road crossing decision (whose most important examples are perceptual, cognitive and motor abilities) has demonstrated that elderly pedestrians are more likely to die or be seriously injured in road traffic collisions than adult people \cite{asher2012most}. Elderly pedestrians are indeed strongly conditioned by: (\emph{i}) limitations in locomotion behaviour \cite{winogrond1981comparison} (e.g., reduced range of motion, loss of muscle strength and coordination, changes in posture, decreased walking speed); (\emph{ii}) the progressive decline in the operation of perceptive sensors and cognitive skills \cite{webb2003influence,lobjois2007age} (e.g., limited perception of light and colours, inability to tune out background noise, diminished attention and reaction time, spatial disorientation, slower decision-making). 
  
In this framework, the paper presents the results achieved by means of a video-recorded observation of pedestrian/vehicle interactions at an urban unsignalized intersection in the city of Milan (ITALY), characterised by a significant presence of elderly inhabitants and risky pedestrian-vehicle interactions. 
The methodology which sets the current work is presented in Section 2, with reference to tools used for data collection and the description of data analysis procedures. The results of the observation regarding vehicular and pedestrian traffic volumes and Level of Service are presented in Section 3. The results of the analysis on the time series of the speeds of both adult and elderlies pedestrians are presented in Section 4, with reference to the definition of crossing behaviour as characterised by three sequential phases: approaching, appraising and crossing. The results about crossing trajectories, crossing phases localisation on side-walks and decision making process about the accepted distance and speed of oncoming vehicles are shown in Section 5. The paper concludes with remarks about the achieved results and their future use as supports for the development and calibration of an integrated simulation tool for the representation of pedestrian/vehicles interaction in urban contexts~\cite{DBLP:conf/acri/CrocianiV14}. Future works based on ongoing data collection campaigns are presented at the end of the paper.

\section{Data Collection and Analysis}
\label{sec:method}

\begin{figure}[t!]
\begin{center}
\subfigure{\includegraphics[width=.75\textwidth]{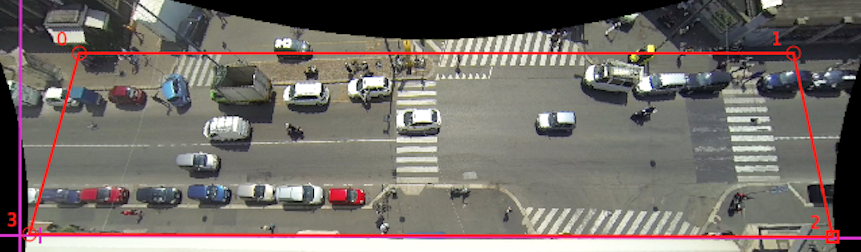}}
\subfigure{\includegraphics[width=.209\textwidth]{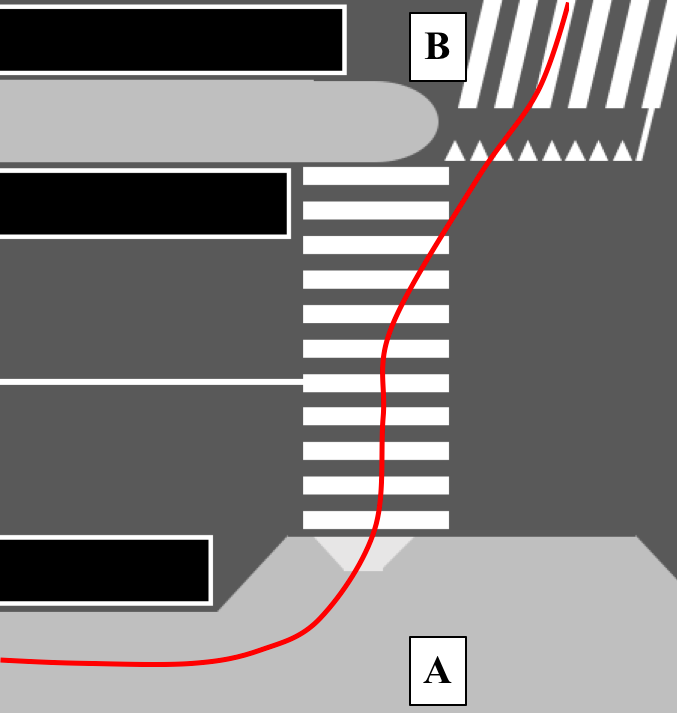}}
\caption{A video frame (a) and a scaled reproduction of the observed zebra crossing with an example trajectory (b).}
\label{fig:scenario}
\end{center}
\end{figure}

The video-recorded observation was performed on May 18, 2015 (from 10:45 am to 12 am), at an urban unsignalized intersection in the city of Milan (ITALY). The scenario of the observation (the intersection between Via Padova, Via Cambini and Via Cavezzali) has been selected by means of a preliminary analysis which was aimed at crossing the geo-referred information related to the socio-demographic characteristics of the inhabitants of Milan and the localisation of road traffic accidents. Results showed that the chosen residential area is characterised by a significant presence of elderly inhabitants and an high number of pedestrian/car accidents involving elderlies pedestrians in the past years\footnote{See \url{http://aim.milano.it/en/pubblicazioni-en/archivio-pubblicazioni-en}}.

A series of inspections of the residential area allowed to select that particular unsignalized intersection among others, considering the large amount of people which pass through it due to several points of interest (e.g., local market, public offices, bank, supermarket, Church, Islamic cultural centre). The observation was performed during the peak hour of the open-air local market which is held every Monday in Via Cambini (nearby point B). Weather conditions during the observation were stable and sunny.

A HD ultra wide lens camera was mounted on a light stand tripod overhung from the balcony of a private flat in correspondence of the zebra crossing (at an height of about 25 m). The hidden position of the camera allowed to not influence the behaviour of drivers and pedestrians. Some spatial reference points have been taken to correct the distortion of video frames due to the wide lens and the nearly zenith perspective of the camera (see Fig. \ref{fig:scenario}/a).

A first phase of data analysis consisted of manual counting activities to estimate and characterise the traffic volumes (1379 vehicles, 585 crossing pedestrians). An \emph{ad hoc} checklist comprising a set of locomotion and physical indicators was used to support the annotators in profiling pedestrians\rq\ age and gender, and the presence of pedestrians crossing in groups. Then, a series of time stamping activities were aimed at measuring the additional travel time experienced by drivers and pedestrians due to traffic conditions, in order to determinate the Level of Service of the observed intersection. 

A second phase of data analysis was based on the use of the open source software Tracker Video Analysis and Modelling Tool\footnote{See \url{www.cabrillo.edu}} (see Fig. \ref{fig:scenario}), which allowed to correct the distortion of the video images, and then to semi-automatically track a sample of 50 pedestrians and 79 vehicles while interacting at the zebra crossing, considering one frame every ten (every 0.4 sec). The tracking analysis was executed by pointing at the frontal plates of vehicles and at the space in between the feet of pedestrians. The data set (including the \emph{X}, \emph{Y} coordinates and the associated frames \emph{t}) was exported for data analysis, aiming at measuring the speeds of vehicles and pedestrians while interacting at the zebra crossing and the safety gap accepted by pedestrians to cross, comparing data among adults and elderlies. 

\section{Traffic Volumes, Pedestrians Profiling and Level of Service}
\label{sec:traffic_pedestrians}

The bidirectional flows of vehicles and pedestrians passing through the considered crosswalk have been counted minute by minute and classified with reference to their typologies and profiles. Vehicular traffic volumes were constituted for the large majority by cars (67\%); the direction of movements of vehicles was equally distributed. Adult singles were the majority of the total counted pedestrians (71\%); elderlies were a significant portion of the pedestrians flows (24\%). The 90\% of the total pedestrian flows on the zebra crossing was directed towards the local market area or was oncoming from it.

\begin{itemize}[noitemsep,leftmargin=*]
\item Total: 1379 vehicles, 18.89 vehicles per minute;
\item Type: 922 (67\%) cars, 181 (13\%) bikes, 115 (8\%) vans, 82 (6\%) cycles, 43 (3\%) buses, 36 (3\%) trucks;
\item Direction: 685 (50\%) vehicles toward the city centre; 693 (50\%) vehicles toward the neighbours.\\
\item Total: 585 pedestrians, 8.01 pedestrians per minute;
\item Age: 415 (71\%) adults, 138 (24\%) elderly (from about 65 y.o.), 32 (5\%) young (until about 18 y.o.);
\item Gender: 307 (52\%) males, 278 (48\%) females;
\item Grouping: 381 (65\%) singles, 76 (26\%) dyads, 16 (8\%) triples, 1 (1\%) group $\geq$ four members;
\item Direction: 340 (58\%) pedestrians toward the local market; 245 (42\%) ped. oncoming from local market.
\end{itemize}

The Level of Service (LOS) \cite{milazzo1999quality} standardly describe the degree of comfort and safety afforded to drivers and pedestrians as they travel/walk through an intersection or roadway segment. Six grades are used to denote the various LOS from ÒAÓ to ÒFÓ, by measuring the additional travel time (delay) experienced by drivers and pedestrians, as an important indicator of the efficiency of an intersection. At two-way stop-controlled unsignalized intersections (unsignalized zebra crossings in which pedestrians have the right-of-way) LOS ÒEÓ is considered to represent the minimum acceptable design standard (see Tab. \ref{tab:los}). 

The LOS have been estimated by time stamping the delay of vehicles due to vehicular and pedestrian traffic conditions (time for deceleration, queue, stopped delay, acceleration), and the delay of crossing pedestrians due to drivers\rq\ non compliance to pedestrian right of way (waiting, start-up delay). Results showed that both the average delay of vehicles (3.20 s/vehicle $\pm$ 2.73 sd) and the average delay of pedestrians (1.29 s/pedestrian $\pm$ .21 sd) corresponded to LOS A. 

According to the recommendations of \cite{milazzo1999quality}, the measured traffic volumes (1139 vehicles/hour/both directions) are sufficiently high to envisage the implementation of a traffic light system (although more thorough observations in other times/weather conditions should be performed to actually reach an informed decision). Results about LOS showed that nearly all drivers found freedom of operation and that no pedestrians crossed irregularly, with low risk-taking crossing behaviour. 

\begin{table}[t!]
\small
\centering
\caption{The Level of Service criteria for two-way stop-controlled unsignalized intersections \cite{milazzo1999quality}.}\vspace{0.5cm}
\label{tab:los}
\begin{tabular}{|c|p{8.8cm}|c|c|}
\hline
\cellcolor[HTML]{EFEFEF}                      & \cellcolor[HTML]{EFEFEF}                                                                                                                                                                                                                                                                                                                                                            & \cellcolor[HTML]{EFEFEF}                                                                                                 & \cellcolor[HTML]{EFEFEF}                                                                                                       \\
\multirow{-2}{*}{\cellcolor[HTML]{EFEFEF}LOS} & \multirow{-2}{*}{\cellcolor[HTML]{EFEFEF}Description}                                                                                                                                                                                                                                                                                                              & \multirow{-2}{*}{\cellcolor[HTML]{EFEFEF}\begin{tabular}[c]{@{}c@{}}Vehicular Delay\\ {[}s/veh{]}\end{tabular}} & \multirow{-2}{*}{\cellcolor[HTML]{EFEFEF}\begin{tabular}[c]{@{}c@{}}Pedestrian Delay\\ {[}s/ped{]}\end{tabular}} \\ \hline
\cellcolor[HTML]{007FFF}A                     & \begin{tabular}[c]{@{}l@{}}- Nearly all drivers find freedom of operation\\ - Very small delay, no one cross irregularly\end{tabular}                                                                                                                                                                                    & $<$ 5                                                                                                              & $<$ 10                                                                                                                   \\ \hline
\cellcolor[HTML]{34CDF9}B                                             & \begin{tabular}[c]{@{}l@{}}- Occasionally there is more than one vehicle in queue\\ - Small delay, almost no one cross irregularly\end{tabular}                                                                                                                                                                & 5 - 10                                                                                                                   & 10 - 15                                                                                                                        \\ \hline
\cellcolor[HTML]{32CB00}C                                             & \begin{tabular}[c]{@{}l@{}}- Many times there is more than one vehicle in queue\\ - Small delay, very few pedestrian cross irregularly\end{tabular}                                                                                                                                                            & 10 - 20                                                                                                                  & 15 - 25                                                                                                                        \\ \hline
\cellcolor[HTML]{F8FF00}D                                             & \begin{tabular}[c]{@{}l@{}}- Often there is more than one vehicle in queue\\ - Big delay, some ped. start crossing irregularly\end{tabular}                                                                                                                                                                                                    & 20 - 30                                                                                                                  & 25 - 35                                                                                                                        \\ \hline
\cellcolor[HTML]{F8A102}E                                             & \begin{tabular}[c]{@{}l@{}}- Drivers find the delays approaching intolerable levels\\ - Very big delay, many pedestrians cross irregularly\end{tabular} & 30 - 45                                                                                                                  & 35 - 50                                                                                                                        \\ \hline
\cellcolor[HTML]{FE0000}F                                             & \begin{tabular}[c]{@{}l@{}}- Forced flow due external operational constraints \\ - Pedestrians cross irregularly, engaging risk-taking behaviours\end{tabular}                                                                        & $>$ 45                                                                                                          & $>$ 50                                                                                                                \\ \hline
\end{tabular}
\end{table}

\section{Trend Analysis of Pedestrian Speeds and Crossing Phases}

A sample of 50 pedestrians and 79 vehicles was considered for data analysis. The sample was selected avoiding situations such as: platooning of vehicles on the roadway inhibiting a crossing episode, the joining of pedestrians already crossing, and in general situations influencing the direct interaction between the pedestrians and the drivers. Part of the selected crossing episodes was characterised by the multiple interaction between the crossing pedestrian and two vehicles oncoming from the near and the far lane. Considering the effects of several interfering variables on results, the sample has been designed as follows: 

\begin{itemize}[noitemsep,leftmargin=*]
\item Age: 27 adult and 23 elderly crossing pedestrians;
\item Direction: 27 pedestrians toward point B; 23 pedestrians toward point A;
\item Gender: 22 males, 28 females;
\item Lane: 50 vehicles from the near lanes, 29 vehicles from the far lane;
\item Vehicle typologies: 77 cars, 1 motorbike, 1 bus.
\end{itemize}

\begin{figure}[t!]
\begin{center}
\subfigure{\includegraphics[width=.7\textwidth]{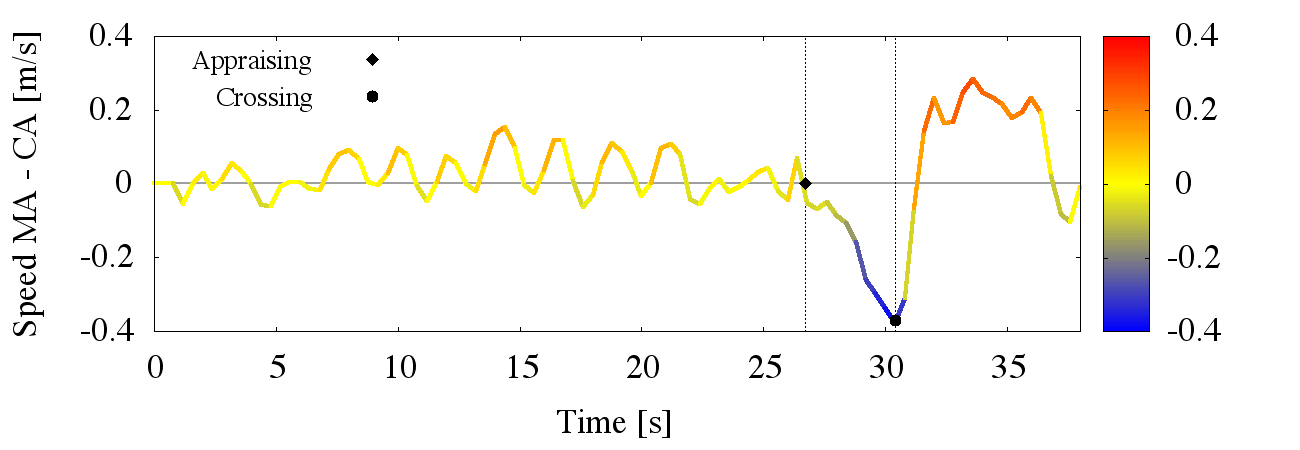}}
\caption{An exemplification of the trend analysis performed on the time series of speeds.}
\label{fig:phases}
\end{center}
\end{figure}

Pedestrian speeds have been analysed among the time series of video frames (trend analysis), as characterised by: (i) a stable trend on side-walks, (ii) a significant deceleration in proximity of the cross-walk (decision making) and (iii) an acceleration on the zebra crossing. The trend of speeds was analysed by calculating the difference between: the moving average (MA, time period length: 0.8 s, three frames), and the cumulative average (CA) of the entire frames series. This allowed to smooth out short-term fluctuations of data (intrinsically due to pedestrian gait, but also caused by the frame discretization) and to highlight longer-term trends (deceleration/acceleration). According to results, crossing behaviour is defined as composed of three distinctive phases (see Figure \ref{fig:phases}):

\begin{enumerate}[label=\alph*),noitemsep, leftmargin=*]
\item Approaching: the pedestrian travels on the side-walk with a stable speed (Speed MA - CA $\simeq$ 0);
\item Appraising: the pedestrian approaching the cross-walk decelerates to evaluate the distance and speed of oncoming vehicles (safety gap). We decided to consider that this phase starts with the first value of a long-term deceleration trend (Speed MA - CA $<$ 0);
\item Crossing: the pedestrian decides to cross and speed up. The crossing phase starts from the frame after the one with the lowest value of speed before a long-term acceleration trend (Speed min).
\end{enumerate}

\begin{figure}[t!]
\begin{center}
\subfigure[Pedestrians crossing from A to B]{\includegraphics[width=.48\textwidth]{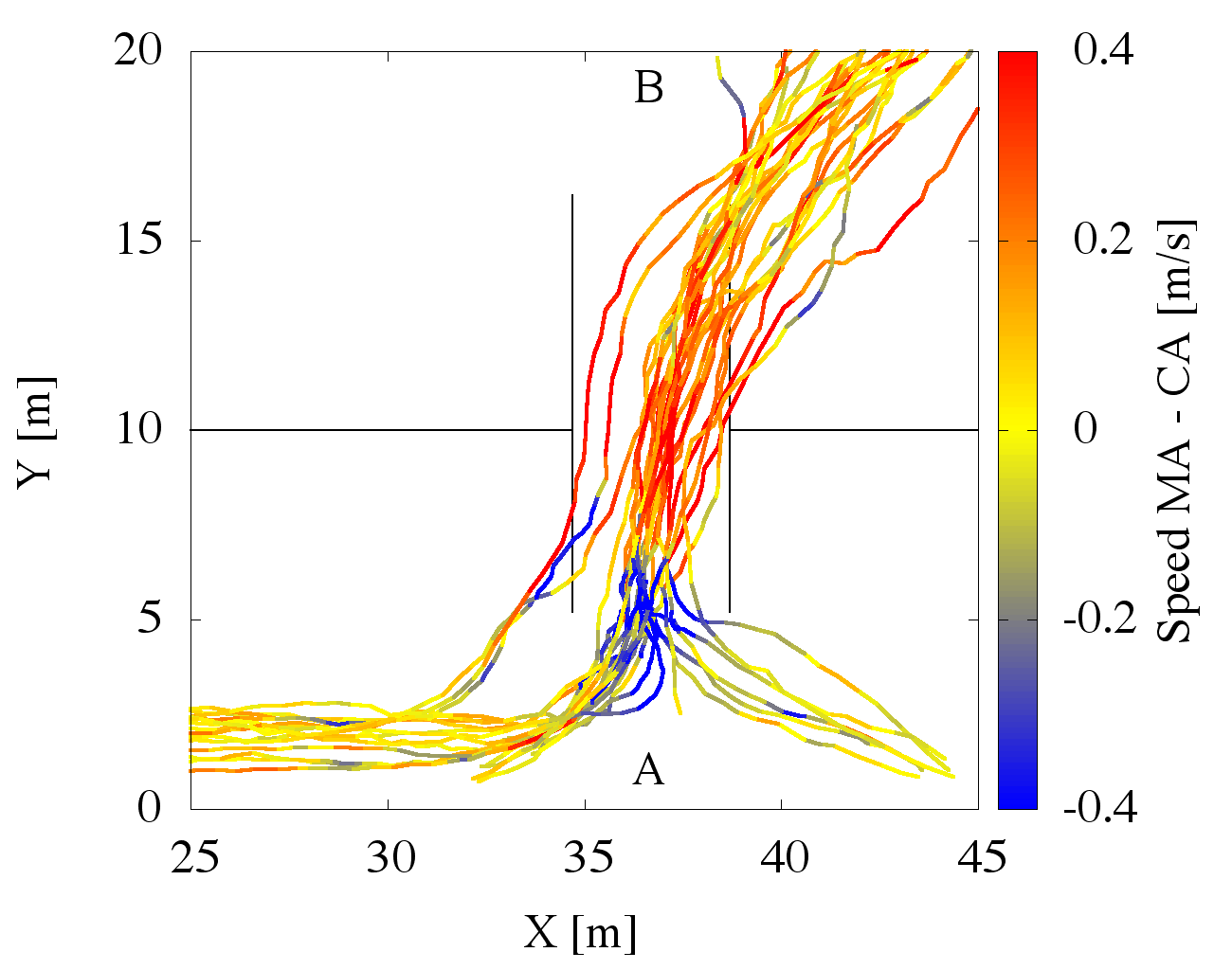}}\hspace{.5cm}
\subfigure[Pedestrians crossing from B to A]{\includegraphics[width=.48\textwidth]{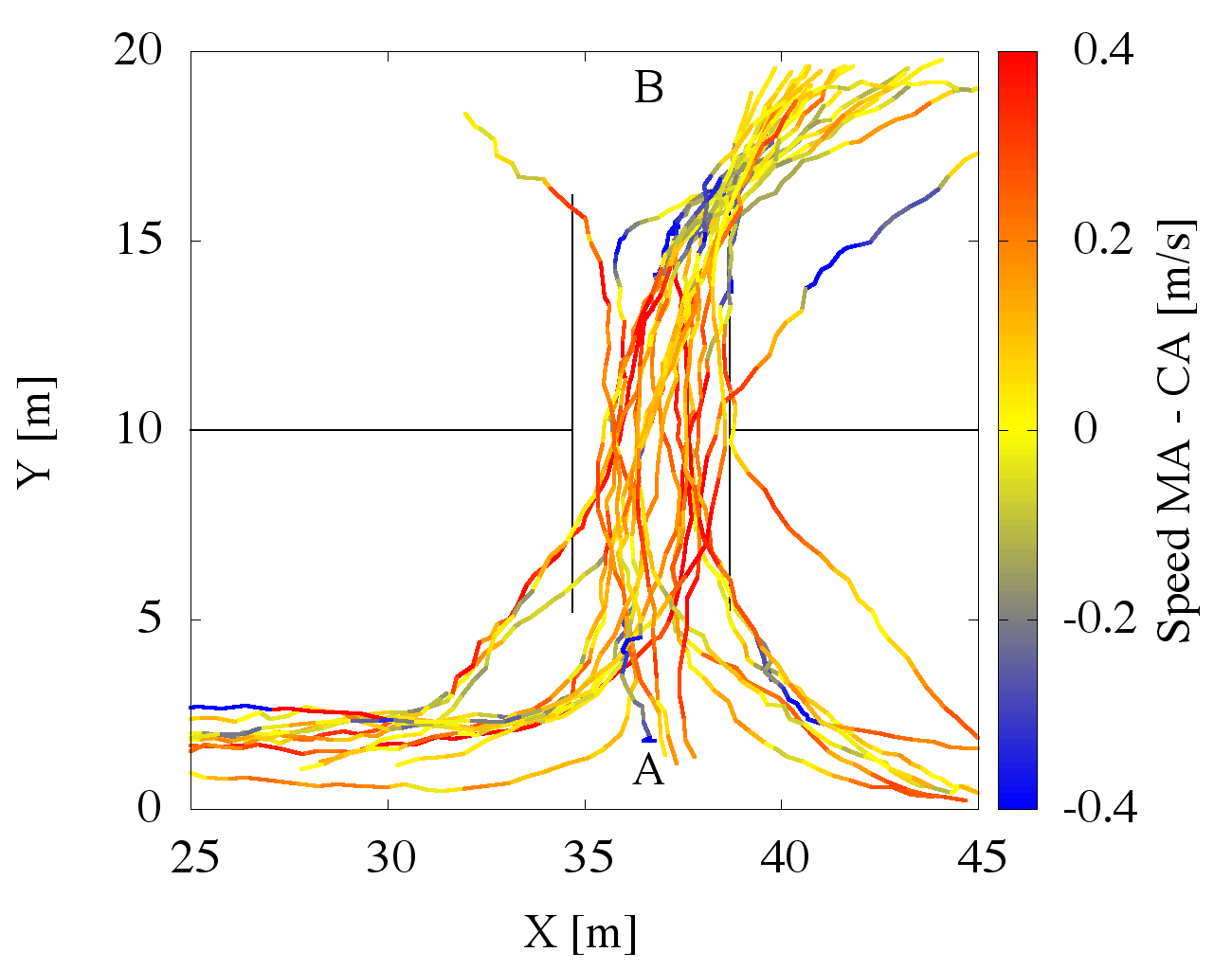}}
\caption{The trajectories of crossing pedestrians plotted with the results of the trend analysis of speeds, which highlighted the difference of results among the crossing phases. The charts report the central line of the road way and spatial references of the zebra-striped crossing.}
\label{fig:tr}
\end{center}
\end{figure}

A two-factors analysis of variance\footnote{All statistics have been conducted at the p $<$ .01 level.} (two-way ANOVA) showed a significant difference among the speeds of pedestrians while approaching, appraising and crossing [F(2,144) = 61.944, p $=$ .000], and a significant effect of pedestrian\rq\ age on results [F(1,144) = 63.751, p $=$ .000] (see Tab. \ref{tab:speed} and Fig. \ref{fig:tr}). A series of post hoc Tukey test showed a non significant difference between the speeds of pedestrians while approaching and crossing, considering both adults and elderlies (p $>$ .05). The difference between the speed of adults and elderlies was significant among all the three crossing phases (p $=$ .000).

To test the effect of density on speeds \cite{milazzo1999quality}, we measured the flow rate on sidewalks (2.90 ped/min/m $\pm$ .44 sd) and zebra crossing (2.00 ped/min/m $\pm$ .43 sd), which corresponded to a Walkaway LOS A (free flows at low density). Results showed that there was no impact of density on results.

In conclusion, results demonstrated that pedestrians\rq\ crossing decision is based on a significant deceleration in proximity of the curb (appraising) to evaluate the distance and speed of oncoming vehicles. Elderlies walked in average 22\% slower than adults among the three crossing phases, decelerating 6\% more than adults while appraising. This demonstrated the negative impact of ageing on crossing behaviour in terms of locomotion skills decline.

\begin{table}[t!]
\centering
\small
\caption{The speed of adult and elderly pedestrians among the crossing phases.}\vspace{0.5cm}
\label{tab:speed}
\begin{tabular}{|l|l|l|}
\hline
\rowcolor[HTML]{EFEFEF} 
Crossing Phases      &  Adult pedestrians     & Elderly pedestrians   \\ \hline
Approaching speed & 1.28 m/s $\pm$ .18 sd & 1.03 m/s $\pm$ .18 sd \\ \hline
Appraising speed & .94 m/s $\pm$ .21 sd  & .69 m/s $\pm$ .23 sd  \\ \hline
Crossing speed    & 1.35 m/s $\pm$ .18 sd & 1.09 m/s $\pm$ .17 sd \\ \hline
\end{tabular}
\end{table}

\section{Crossing Trajectories, Phases Localisation and Safety Gap}

\begin{figure}[t!]
\begin{center}
\subfigure[Pedestrians/vehicles interaction at point A]{\includegraphics[width=.9\textwidth]{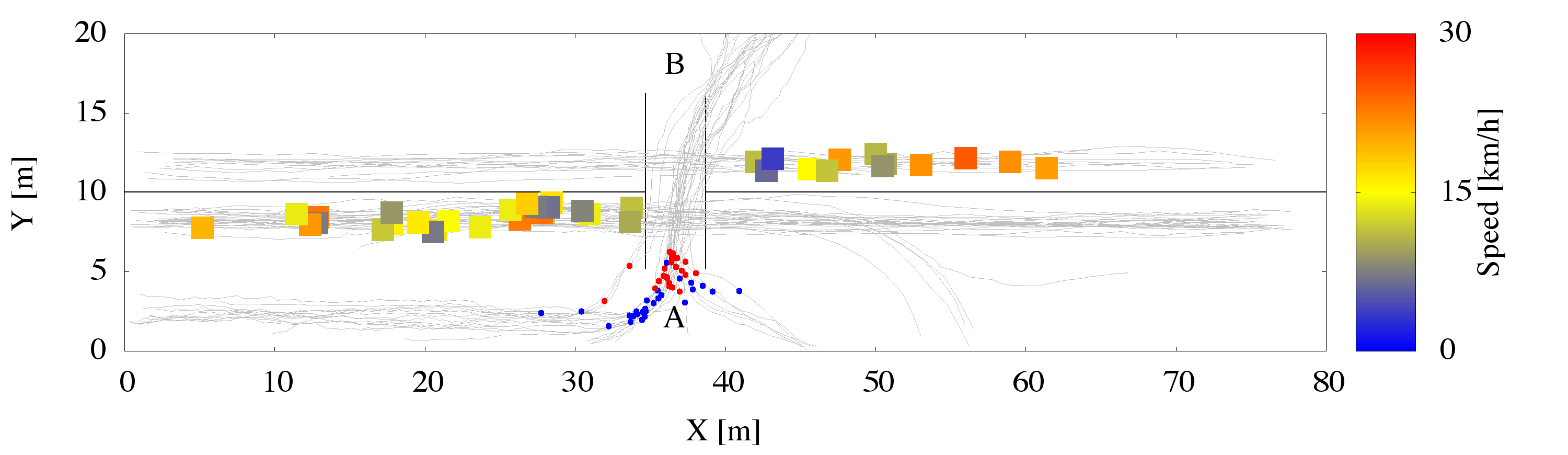}}
\subfigure[Pedestrians/vehicles interaction at point B]{\includegraphics[width=.9\textwidth]{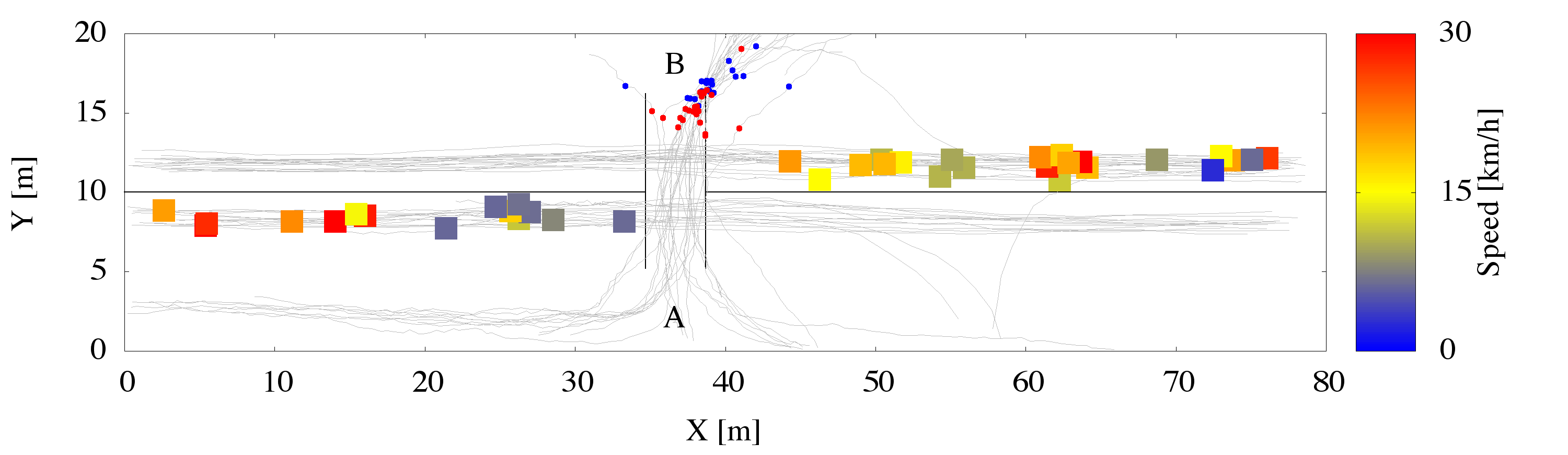}}
\caption{The positions of pedestrians when they started appraising (blue dots) and crossing (red ones). Squared dots refer to the position of vehicles on the near and far lanes, and their speeds. The charts report the central line of the road way and spatial references of the zebra-striped crossing.}
\label{fig:gap}
\end{center}
\end{figure}

Data analysis showed that the average length of trajectories (12.05 m $\pm$ .58 SD) was 9\% longer compared to the shortest path between the curb of points A and point B (11 m). Instead of walking straight, pedestrians cut off the zebra-striped crossing with an oblique path which made them partially occupying the road way (see Fig. \ref{fig:gap}). A series of independent sample t-test analyses showed that the direction of movement and the age of pedestrians did not significantly impact results (p $>$ .05).

The positions of pedestrians when they started appraising and crossing were analysed to estimate their distance from the curb (see Fig. \ref{fig:gap}). At point A, pedestrians started appraising 2.85 m ($\pm$ 1.45 SD) before the curb and they started crossing 0.12 m ($\pm$ 1.02 SD) before the curb. At point B, they started appraising 0.89 m ($\pm$ 1.19 SD) before the curb, but they started crossing 1.17 m ($\pm$ 1.51 SD) after it, occupying the road way. A series of independent samples t-tests showed that the age of pedestrians did not significantly impact results, considering both crossing point A and B (p $>$ .05).

Point A is characterised by a large side-walk area, which allowed pedestrians to safely evaluate the distance and speed of oncoming vehicles; moreover, the presence of a ramp (primarily aimed at facilitating the access to the side-walk for those people with limited mobility) was found to limit non compliant crossing behaviour by gathering pedestrians to the zebra crossing. On the contrary, the small and interrupted side-walk at point B and the presence of various obstacles (e.g., traffic signal pals, taxi station telephone exchange) impeded pedestrians to cross in a safe manner, making them occupying the road way with potentially risky pedestrian-vehicular interactions (see Fig. \ref{fig:scenario}/b).
 
In this paper we denote the \emph{safety gap} as the pedestrians\rq\ evaluation of the distance of an approaching vehicle (measured on the X-axis) and its average speed (not taking into account acceleration/deceleration trends) in order to decide if the gap is large enough to pass safely (see Tab. \ref{tab:gap} and Fig. \ref{fig:gap}). A series of two-way ANOVA showed that, while evaluating the accepted distance and speed of oncoming vehicles, pedestrians did not  discriminate between the near and far lanes (p $>$ .05). Moreover, the age of pedestrians did not impact results (p $>$ .05). An independent samples t-tests showed that evaluating the gap acceptance in a situation comprising one vehicle approaching from the near lane or two vehicles from respectively the near and far lane did not present a significant difference (p $>$ .05). A series of Pearson product-moment correlation coefficient showed that there was a correlation between the speed of oncoming vehicles and their distance; in particular, appraising pedestrians were able to correctly evaluate these variables both considering the near lane [r = .396, n = 48, p $=$ .005] and the far lane [r = .772, n = 29, p $=$ .000], and this holds for adults [r = .390, n = 43, p $=$ .010] and elderlies [r = .706, n = 34, p $=$ .000].

Results showed that crossing decision making was not influenced by the lane occupied by the oncoming vehicles. Moreover, the time needed by pedestrians to evaluate the safety gap from only one vehicle and two vehicles did not differ significantly. Adults and elderly pedestrians accepted the same distance and speed to safely cross and they were both able to efficiently evaluate the safety gap from vehicles, considering the demonstrated correlation between the estimated distance and speed of approaching vehicles (correctly evaluating the time to arrive of vehicles). However, an independent samples t-test analysis showed a significant difference between the time duration of the appraising phase among elderlies (4.12 s $\pm$ 2.54 sd) and adults (2.79 s $\pm$ 1.47 sd), t(48) = 2.40658, p = .012. Although pedestrians have the right of way on zebra-striped, elderly pedestrians were found to be more cautious than adults: 57\% of them gave way to at least one approaching vehicle, compared to 30\% of adults. This demonstrated the negative impact of ageing on crossing behaviour in terms decision making process about the safety gap.

\begin{table}[t!]
\centering
\small
\caption{The distance and speed of vehicles from the near and far lanes, considering adults and elderlies}\vspace{.5cm}
\label{tab:gap}
\begin{tabular}{|l|l|l|}
\hline
\rowcolor[HTML]{EFEFEF} 
Vehicles from the near lane & Adult crossing pedestrians        & Elderly crossing pedestrians      \\ \hline
Accepted Distance from vehicles          & 15.68 m $\pm$ 8.94 sd    & 20.16 m $\pm$ 10.30 SD   \\ \hline
Accepted Speed of vehicles            & 16.54 km/h $\pm$ 5.91 SD & 14.82 km/h $\pm$ 7.28 SD \\ \hline
\rowcolor[HTML]{EFEFEF} 
Vehicles from the far lane  & Adult pedestrians        & Elderly pedestrians      \\ \hline
Distance from vehicles                   & 17.12 m $\pm$ 6.56 SD    & 16.11 m $\pm$ 10.12 SD   \\ \hline
Speed of vehicles             & 16.53 km/h $\pm$ 7.57 SD & 14.10 km/h $\pm$ 9.13 SD \\ \hline
\end{tabular}
\end{table}

\section{Conclusions and Future Works}

The issue of developing urban sustainable transportation strategies is becoming a mandatory requirement for municipalities to enhance the quality of life and the safety of the citizens. In particular, the social cost of pedestrians\rq\ risky behaviour pushes the development of a new generation of computational models integrating analytical knowledge, data and experience about the complex dynamics occurring in pedestrian/vehicle interactions, supporting urban and traffic decision makers and managers. 

The current work presented the results of an empirical study focused on pedestrian crossing behaviour in an urban unsignalized intersection in the city of Milan (ITALY), characterised by a significant presence of elderly inhabitants and risky pedestrian-vehicle interactions. The achieved results are finally aimed at supporting the development and the parametric calibration of advanced simulation systems, focusing on specific behavioural rules regulating pedestrians\rq\ crossing decision making while interacting with approaching vehicles at unsignalized intersections \cite{DBLP:conf/acri/CrocianiV14}. 

In general, pedestrian crossing behaviour requires the coordination of complex perceptive, attentional and locomotion skills. Several environmental factors could negatively influence pedestrian decision making (e.g., sidewalk discontinuity and condition, lack of curb ramps, obstructions in walkway, limited visibility), with potential risky consequences for their safety \cite{asher2012most}. The result of the current work showed that crossing behaviour is based on three sequential phases (approaching, appraising, crossing), characterised by deceleration/acceleration trends. In particular, pedestrians\rq\ appraising phase is based on a significant deceleration of speeds in proximity of the zebra-striped crossing to evaluate in time the distance and speed of oncoming vehicles (decision making). Results showed that, due to the decline of locomotion and perceptive skills linked to ageing, elderly pedestrians walked significantly slower than adults and they spent more time to evaluate the safety gap from oncoming vehicles. This demonstrated the impact of ageing on crossing behaviour, in terms of locomotion and decision making process. However, results showed that both adults and elderlies correctly evaluated the distance and speeds of the vehicles oncoming from the near and the far lanes, with no risk-taking crossing behaviour. 

Although the results about LOS showed that nearly all drivers found freedom of operation and that no pedestrians crossed irregularly at the observed intersection, the comparison of results between crossing point A and B highlighted that the lack of adequate space and ramp on sidewalk at point B made pedestrian crossing in a unsafe manner, occupying the road way. Once calibrated, a series of analyses on simulation results will allow to test the efficiency of different planning and architectural hypotheses aimed at exploring potential ways to improve the overall safety of the observed zebra crossing.  

Ongoing works are based on the use of questionnaires to assess the perceived \emph{walkability} \cite{abley2005walkability} of the area by the elderly inhabitants (e.g., quality of sidewalks, drivers\rq\ compliance). Moreover, we will further analyse the recorded video frames to empirically investigate the overall drivers\rq\ compliance with crossing pedestrians and speed adaptation of vehicles in proximity of the zebra crossing (drivers\rq\ appraising phase) \cite{varhelyi1998drivers}, as well as non compliant jaywalking behaviour of pedestrians out of zebra-striped crossing. Finally, since the analysis of pedestrian profiles showed that the 35\% of the observed population was composed of walking groups, future works will be aimed at testing the impact of grouping on crossing behaviour, comparing results among single pedestrians and dyads. Several works present in the literature, such as \cite{wagner1981crossing}, highlighted indeed that group members use social information, such as the crossing behaviour of others, to judge the safety gap from oncoming vehicles with potentially dangerous crossing decisions.

In conclusion, the results achieved by means of the presented case study could be of notable interest for those involved in modelling pedestrian crossing decision making at unsignalized intersections for developing integrated computer-based simulation tools. Moreover, the results about the speeds of adult and elderly pedestrians while crossing could be used as threshold values for design purposes of pedestrian traffic lights at crossing points (start up and clearance time), but also as supports for the parametric calibration of automated sensor systems for the development of driver-less robotic systems (e.g., collision avoidance from crossing pedestrians). Future and more extended data collection campaigns could be executed on other urban unsignalized intersections, in order to gather consistent data among critical environmental conditions (e.g., traffic volumes, weather conditions, spatial layout of side-walks). 

\subsection*{Acknowledgement}
The Italian policy was consulted and complied in order to exceed the ethical issues about the privacy of the people recorded without their consent. The authors thank Luca Crociani for his valuable contribution.

%\bibliographystyle{IEEEtran}
%\bibliography{template}


\end{document}